\documentclass[twoside]{ilcws08}
\usepackage[latin1]{inputenc}
\usepackage[dvips]{graphicx,epsfig,color}
\usepackage{wrapfig,rotating}
\usepackage{amssymb,amsmath,array}

\begin{document}
\title{
Alignment of Silicon tracking systems
R\&D on Semitransparent Microstrip Sensors} 
\author{
 Jordi Duarte, Marcos Fernández,
Javier González, Richard Jaramillo, \\
Amparo López,
Celso Martínez, David Moya, Alberto Ruiz
\thanks{Corresponding author: A. Ruiz, Email address:ruiz@ifca.unican.es. All the signing Institutions belong to the Spanish Thematic Network for Future Accelerators.Work supported by the Commission of the European Communities (contract RII3-026126), Ministerio de Ciencia y Tecnología-Spain
(projects FPA2007-66387, FPA2007-29110-E)and Consolider-Ingenio 2010 (project CSD2007-0042)},
Ivan Vila
\vspace{.1cm}\\
Instituto de F\'{i}sica de Cantabria \\
(CSIC-Univ. de Cantabria)\\
39005-SANTANDER (SPAIN)\\
\vspace{.3cm}\\
D.Bassignana, E. Cabruja, M. Lozano, G. Pellegrini
\vspace{.1cm}\\
Centro Nacional de Microelectrónica \\
IMB-CNM/CSIC- Barcelona (Spain)}

\maketitle

\begin{abstract}
We summarize the R\&D activities on a novel semitransparent microstrip sensor to be used on laser-based alignment systems for  silicon trackers. The new sensor is used both for particle tracking and laser detection. The aim of this research line is to increase the optical transmittance (T) of Silicon microstrips detectors to infrared light, introducing minor modifications to the sensor design still suitable for its industrial production. The optical simulations used in the sensor design have been experimentally validated against several patterned material samples. This activities have been carried out in the context of SiLC collaboration for the next International Linear Collider.

\end{abstract}

\section{Introduction}

The tracking precision demanded for the International Linear Collider
implies that any environmental
disturbances around the detector, as local temperature
gradients or humidity changes will induce instability of
the supporting structures comparable to the precision of
the detectors. For that reason, alignment systems monitoring
these changes are needed. Silicon sensors currently used as tracking detectors in high energy physics experiments
have a weak absorption (A) of infrared light (IR), but enough to use laser beams as pseudo-tracks traversing consecutive
sensors.
This technique has been successfully used in AMS
and CMS tracker systems \cite{AMS,CMS}.
Narrowing the width of the Al strips and tuning the thicknesses of the
different layers of the sensor, such that they work as an antireflection
coating for the chosen wavelength, we have proved a further 20\%-30\% increase in the transmittance
can be reached.
A thorough simulation of the propagation
of a light beam through the detector has been studied
(section 2) to characterize the optimal geometric parameters of the different components, including effects of multiple reflections in the sensor
layers and diffraction by the strips. We show some measurements
of material samples produced
by the Instituto de Microelectrónica de Barcelona
(IMB-CNM).

\section{Optical Simulation and Design optimization}
Precise alignment and smallest material budget in the tracking system is crucial for the ILC. We propose a hybrid alignment system using a part of the silicon tracking detectors,with minor modifications aiming to make them highly transparent to infrared light.
As a first approximation we modeled a silicon sensor as a stack of
perfectly homogeneous planoparallel layers characterized by its refraction
index N($\lambda$)= n($\lambda$)+ik($\lambda$) and thickness (d),where the imaginary part is related to the light absorption.

\begin{wrapfigure}{rh}{0.5\columnwidth}
\centerline{\includegraphics[width=0.4\textwidth, height=0.4\textwidth]{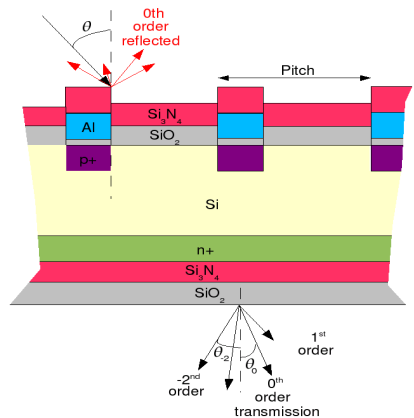}}
\caption{Incoming energy splits in a set of reflection and transmission directions. Represented
only until 2nd order. Shown as well the reference frame, and a
typical slice as used in the simulation.}\label{Fig:Multilayer}
\end{wrapfigure}

In fact the layers are not continuous but present local features (Figure \ref{Fig:Multilayer}),
so diffraction phenomena
will appear if the size of the obstacle is comparable
to the wavelength used.

We have developed a realistic simulation of
the passage of light through a detector,
including interferences effects due to multiple
reflections at the interfaces and diffraction due
to the patterned strips. The aim of the study is to identify key parameters with the goal to optimize the
stack of materials for maximum and robust transparency.

\begin{wrapfigure}{r}{0.4\columnwidth}
\centerline{\includegraphics[scale=0.4]{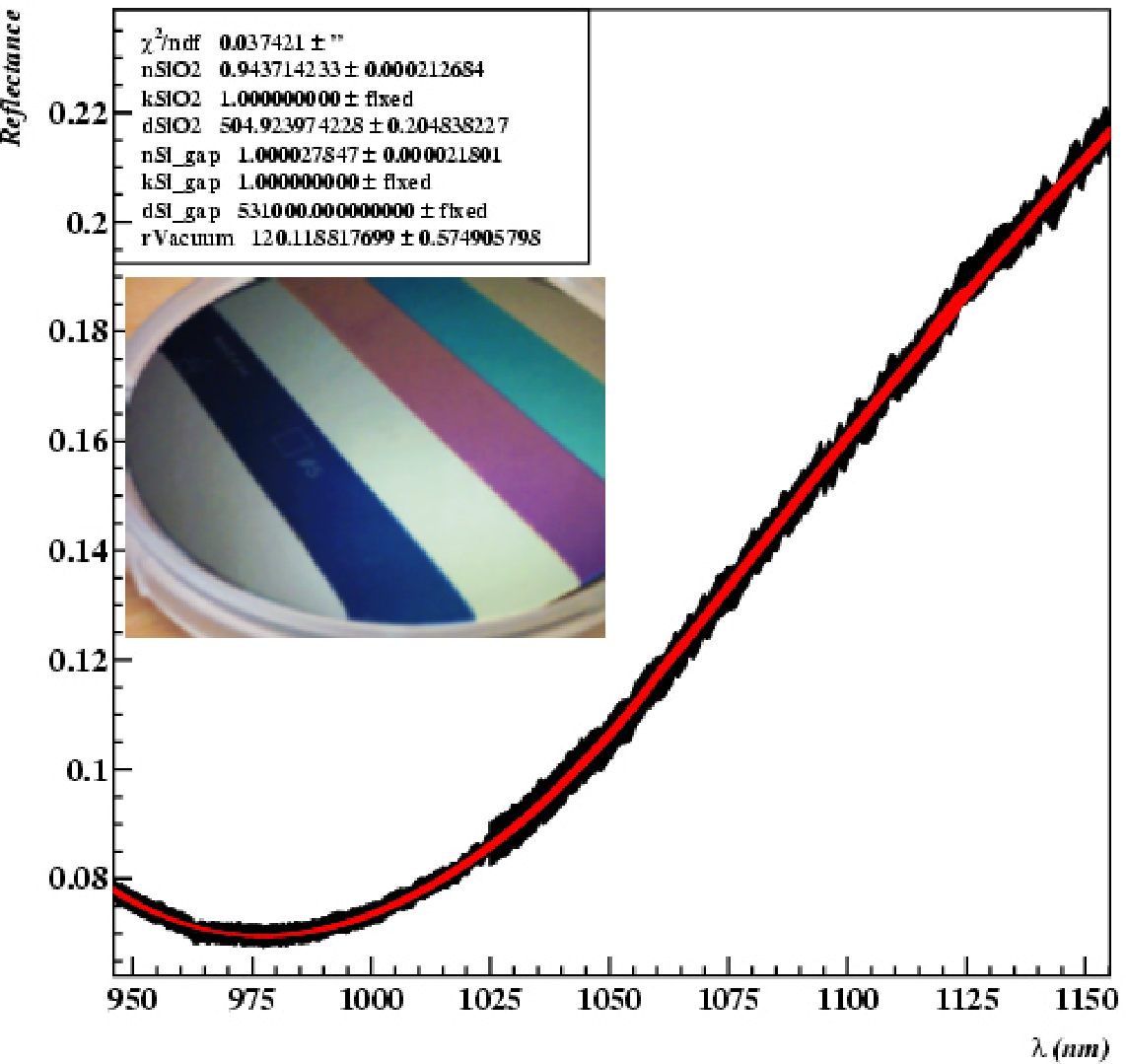}}
\caption{Calculated and measured reflectance of $SiO_2$ on Si.\label{Fig:RSiO2onSi_validation}}
\centerline{\includegraphics[scale=0.25]{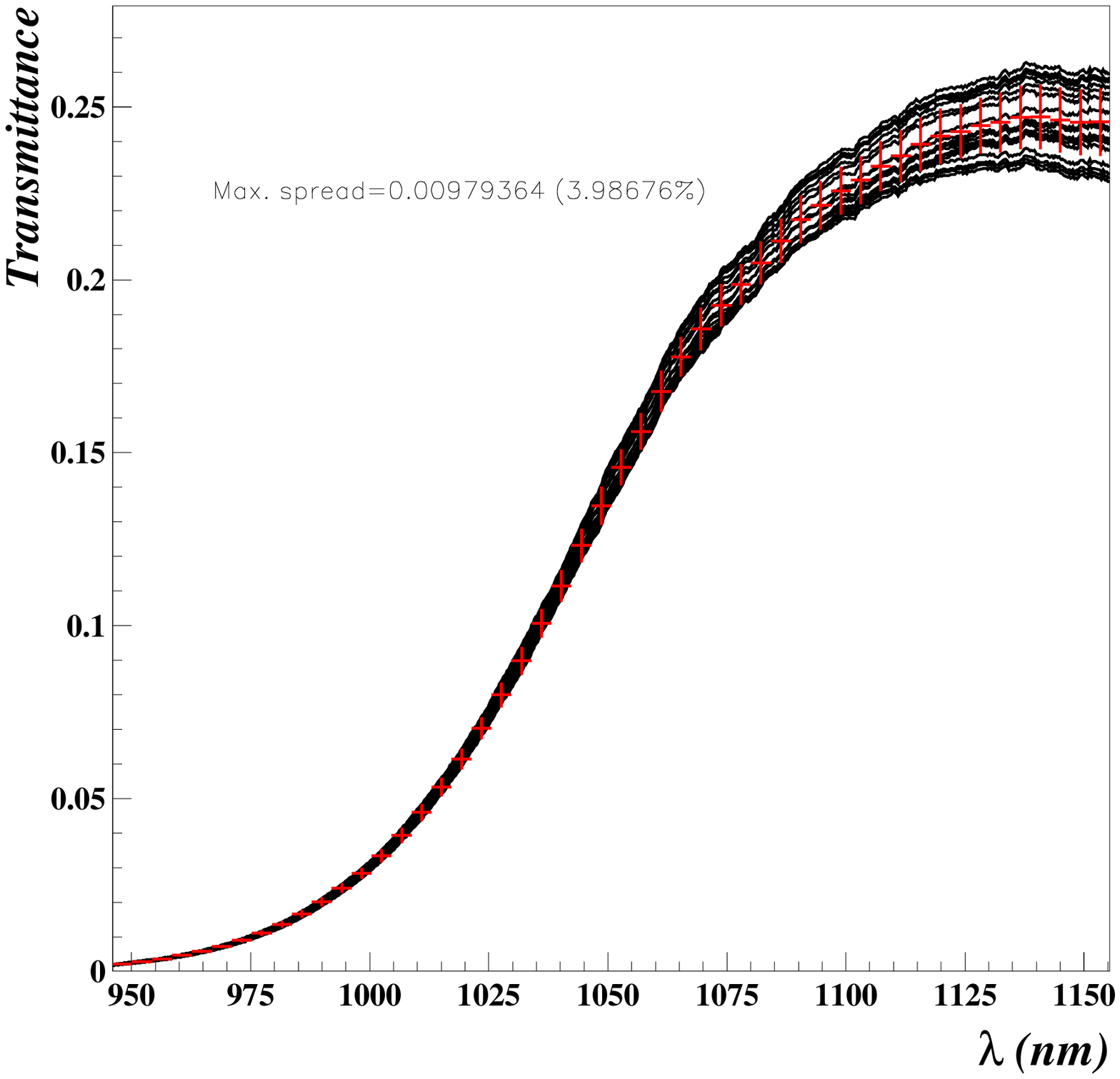}}
\caption{22 measurements(=traces) of T in 3 CMS-like HPK sensors optically treated for alignment.}\label{Fig:T_HPK_centre}
\end{wrapfigure}

The validation of the continuous layer simulation has been made with a
calibrated wafer of SiO2 on Si. A spectrometer with a spectral resolution
of 1.2 nm provides T and reflection (R) as a function of $\lambda$. These measurements are used
to extract the refraction index parameters n($\lambda$), k($\lambda$) and the thickness d,
using the algorithm described in our previous work \cite{EUDET}. For instance, from
the reflectance measurement of the afore-mentioned wafer (see Fig. \ref{Fig:RSiO2onSi_validation}) we
fitted a thickness of SiO2 with only 2 nm error with respect to
the calibrated (known) thickness.

The strips simulation were validated against other authors for both TE and TM polarizations \cite{geist,green} and experimentally validated using photolithographic aluminium patterns on a silicon wafer.
In this mode we can select a wavelength for which the parameterized
transmittance has a relative maximum. Then, we can select the thickness value for each layer i that yields maximum
transmittance, which will display a periodicity with thickness. That will yield a good starting point for
our optimization.
\newpage

Finding an absolute extreme within a range of parameters, requires all the thicknesses
to be changed simultaneously. We built a minimization function, which includes tolerances respect to Montecarlo.\\

\section{Sensor optical characterization}
 We are developing a programme \cite{GICSERV} of sensor optical characterization which includes testing different width/pitch and intermediate strips combinations, polysilicon as strips and doping concentrations. The wafers incorporate optical and electrical test structures.

We have already analyzed some CMS-like Hamamatsu sensors modified to be aligned with lasers. The Al in a circular spot in the back has been removed. A total of 22 measurements made on 3 sensors across the central spot of 1 cm diameter The results are shown in (Figure \ref{Fig:T_HPK_centre}).The uniformity is about 4\%, and T is similar to the CMS alignment sensors. The sensors has been measured with a 1060 nm pulse laser diode and the position of the laser spot has been fitted by a Gaussian fit showing very good performance.

\vskip 2cm


\begin{footnotesize}

\end{footnotesize}


\end{document}